\documentclass[aps,prl,showpacs,notitlepage,floatfix,nofootinbib,preprintnumbers,superscriptaddress,amsmath,amssymb]{revtex4-1}
\usepackage{graphicx}
\usepackage{epsfig}
\usepackage{bm}
\usepackage{color}
\usepackage{float}
\usepackage{dcolumn}

\graphicspath{ {images/} }

\begin{document}

\title{Erratum: The landscape of two-proton radioactivity\\
{[Phys. Rev. lett.  110, 222501 (2013)]}}

\author{E. Olsen, M.~Pf\"utzner, N. Birge, M. Brown, W. Nazarewicz, and A. Perhac}

\maketitle

In our Letter, the proton pairing gaps were incorrectly read from the file containing their calculated values.  This affected our predicted values of $Q_p$ for even-even nuclei and of $Q_p, Q_{2p}$ and $Q_{\alpha}$ values for even-odd systems. Consequently, the text in the \textit{Results} section needs to be corrected together with Figs. 1 and 2.

To identify cases where true, simultaneous (\emph{2p}) emission is the dominating decay mode, following
a discussion in Ref.~[1], we apply the energy criterion $Q_{2p} > 0, Q_p < 0.2 \, Q_{2p}$ 
which is less restrictive than (1). We find candidates for this type of decay, fulfilling in
addition the half-life constrains (2) and (3) only in elements up to tellurium (see the corrected Fig.~2).

In the region between tellurium and lead, the half-life criteria are found to be fulfilled
for cases which satisfy the energy conditions $Q_{2p} > 0, Q_{2p} > Q_p > 0.2 \, Q_{2p}$ which
indicates the \emph{sequential emission} of two protons (\emph{pp}). 
We do find candidates for this kind of decay
in every even-$Z$ isotope above Te, except in xenon, where alpha decay dominates. 
The predicted average path of sequential $pp$ radioactivity, calculated with the direct model
and shown in the new Fig. 1, practically coincides with the $T_{2p}=0.1$\,s limit given 
by the diproton model. As stated in our Letter, in nuclei above lead, the $\alpha$-decay mode
is found to be dominating and no measurable candidates for two-proton emission are expected. Consequently, the physics conclusions derived from the corrected Fig.~1 have not changed as compared to the original Fig.~1.

In the region beyond $^{54}$Zn, the 2p-decay candidates which are closest to current
experimental reach and predicted by both the direct and diproton models are $^{57}$Ge (3), $^{62,63}$Se (2,1),
$^{66}$Kr (3), and $^{103}$Te (2), where the numbers in parentheses indicate the corresponding
number of neutrons beyond the most neutron-deficient isotope known to date. All other cases, including
the sequential \emph{pp} emitters, are located by more than 3 neutrons away from the present 
body of known isotopes.
This distance increases with atomic number and reaches 14 neutrons for $^{165}$Pb
which is predicted to be the \emph{pp}-emitting lead isotope closest to the drip line.

In a few cases competition between two-proton emission and $\alpha$ decay is predicted. 
The two best candidates, predicted by at least two mass models, are $^{103}$Te and $^{145}$Hf. 
The nucleus $^{103}$Te appears as one of the most interesting cases in our survey. Two mass models (SV-min and
UNEDF1) predict the competition between $\alpha$ decay and true \emph{2p} radioactivity, 
one model (SLy4) predicts the competition between $\alpha$ decay and sequential \emph{pp} decay,
and one model (SkM*) predicts the dominance of $\alpha$ decay. 
In $^{145}$Hf, $\alpha$ decay is predicted to compete with
sequential \emph{pp} emission. 

\begin{figure}[htb]
  \begin{minipage}[t]{0.45\linewidth}
	  \centering
    \includegraphics[width=\linewidth]{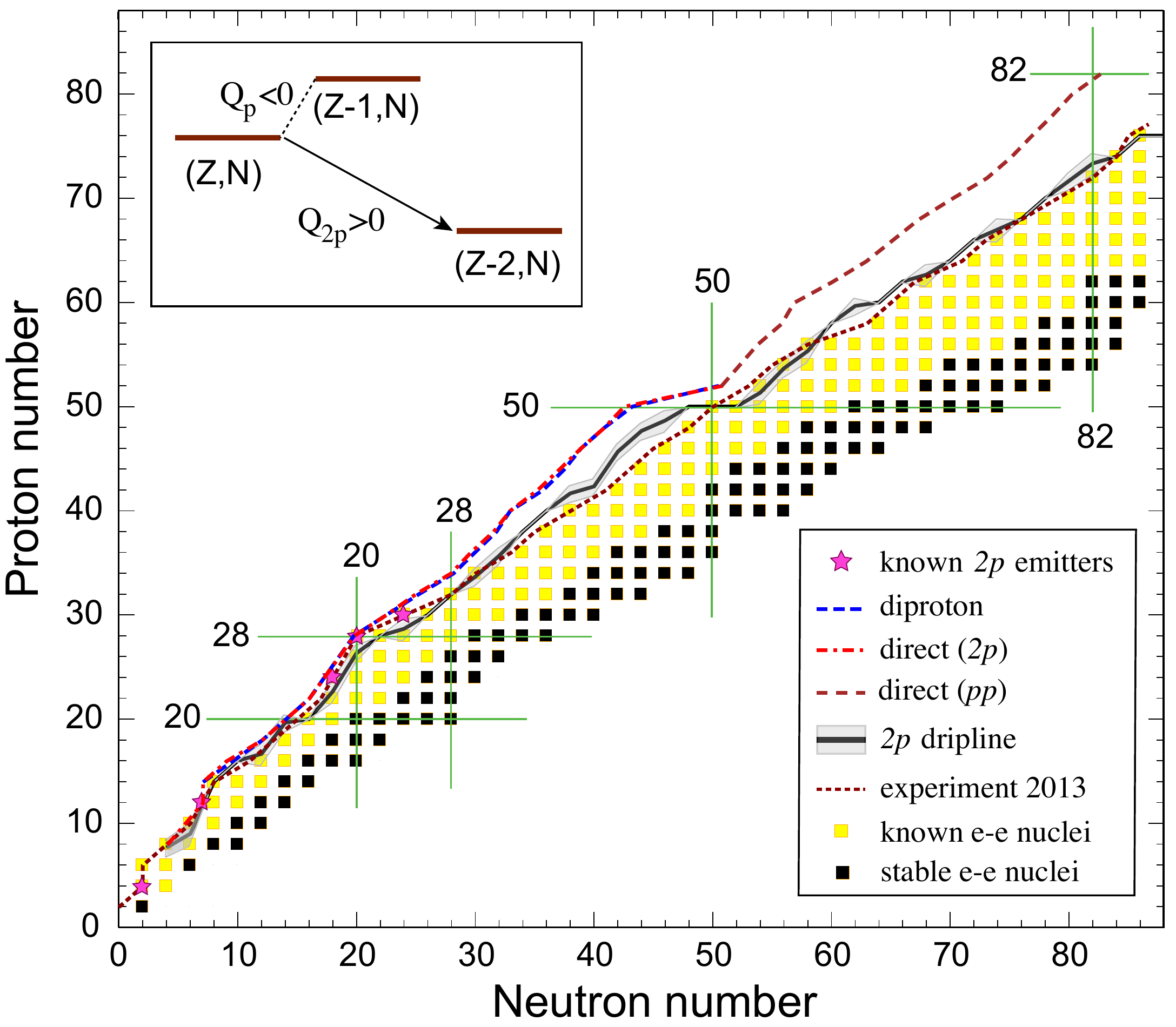}
    \caption{(Color online) Corrected version of Fig. 1. A new line has been added that represents the average path of sequential $pp$ emission as predicted in the direct model. Otherwise the figure caption needs no correction.
}
  \end{minipage}
	\hspace{0.05\linewidth}
	\begin{minipage}[t]{0.45\linewidth}
    \centering
    \includegraphics[width=\linewidth]{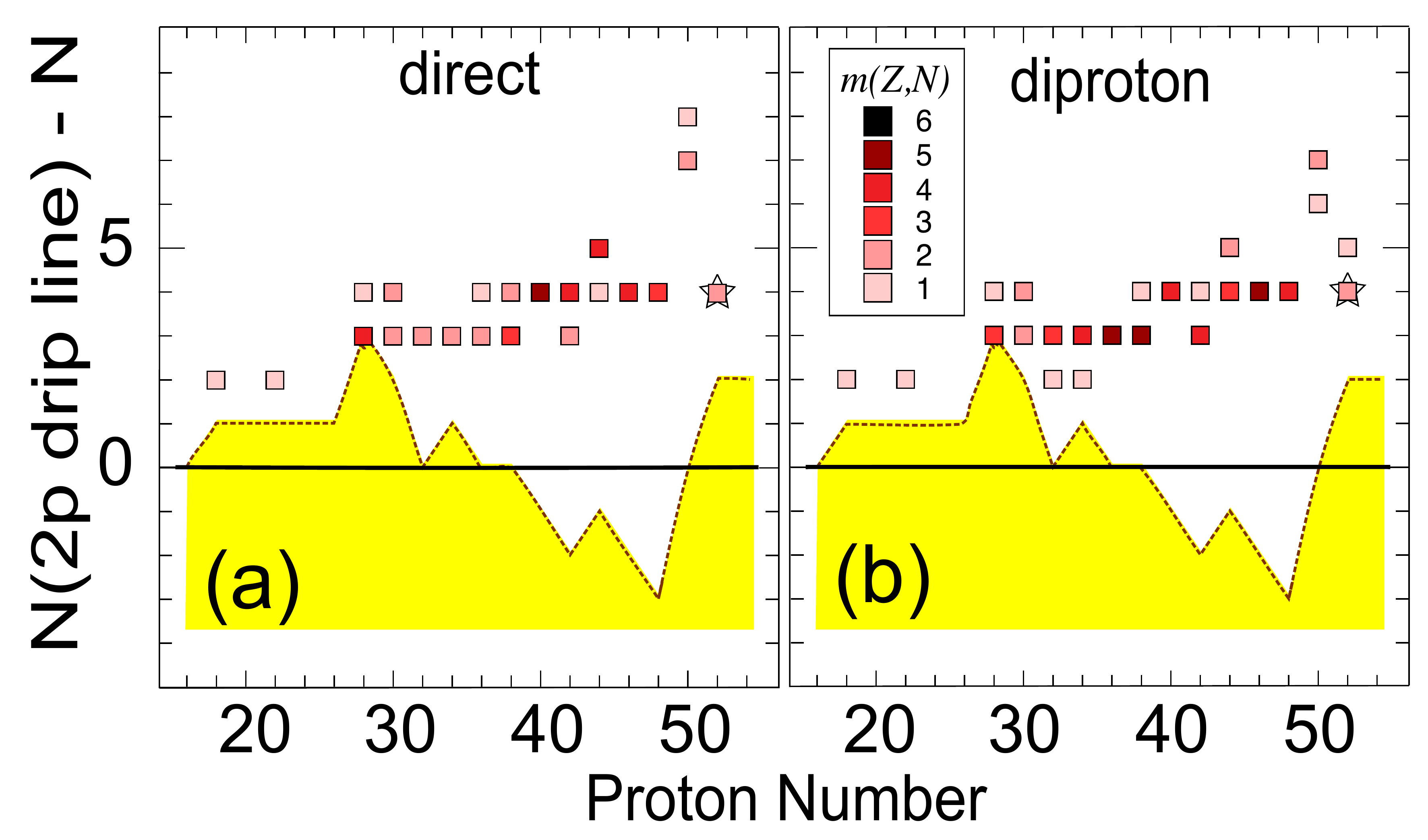}
    \caption{(Color online)  Corrected version of Fig. 2. The figure caption needs no correction.
}
  \end{minipage}
\end{figure}

\end{document}